\documentclass[12pt]{iopart}
\usepackage{graphicx}

\begin{document}

\title[]{Angular distribution measurement of atoms evaporated from a resistive oven applied to ion beam production}

\author{A. Leduc$^{1,2}$, T. Thuillier$^1$, L. Maunoury$^2$, and O. Bajeat$^2$}

\address{$^1$Univ. Grenoble Alpes, CNRS, Grenoble INP\footnote{Institute of Engineering Univ. Grenoble Alpes}, LPSC-IN2P3, 38000 Grenoble, France
\\$^2$GANIL, bd Henri Becquerel, BP 55027,F-14076 Caen, France}
\ead{thuillier@lpsc.in2p3.fr}
\vspace{10pt}
\begin{indented}
\item[]January 2021
\end{indented}

\begin{abstract}
A low temperature oven has been developed to produce calcium 
   beam with Electron Cyclotron Resonance Ion Source (ECRIS). The atom flux from the oven has been studied experimentally as a function of the temperature and the angle of emission by means of a quartz microbalance. The absolute flux measurement permitted to derive Antoine's coefficient for the calcium sample used :  $A=8.98\pm 0.07$ and $B=7787\pm 110$ in standard unit. The angular FWHM of the atom flux distribution is found to be  $53.7\pm7.3$ °at 848K, temperature at which  the gas behaviour is non collisional. The atom flux hysteresis observed experimentally in several laboratories is explained as follows: at first calcium heating, the evaporation comes from the sample only, resulting in a small evaporation rate. once a full calcium layer has formed on the crucible refractory wall, the calcium evaporation surface includes the crucible's enhancing dramatically the evaporation rate for a given temperature.  A Monte-Carlo code, developed to reproduce and investigate the oven behaviour as a function of temperature is presented. A discussion on the gas regime in the oven is proposed as a function of its temperature. A fair agreement between experiment and simulation is found.
\end{abstract}

%
%
%
%
%

\section{Introduction}

This work is dedicated to the study of a low temperature metallic oven dedicated to calcium beam production at the SPIRAL2 facility at GANIL, France~\cite{REF_FOUR_GANIL,REF_PHOENIXV3_2}. The motivation of the study is to better understand the physics and chemistry of such an oven with the goal to improve the global conversion efficiency of rare and expensive isotope  atom (like $^{48}$Ca) to an ion beam in Electron Cyclotron Resonance Ion Source (ECRIS). This study is the first step of longer term plan  to build an end-to-end simulation able to optimize and predict the atom to ion conversion yield of oven to produce beams in ECRIS. Here, the differential atom flux from the oven is measured and compared to simulation. Indeed, the angle of atom emission is an important geometrical factor of the whole atom to ion conversion in an ECRIS. In a first part, the calcium oven used is presented in detail. In a second part, the experimental setup used to study the differential atom flux from the oven is described. Next,  the Monte Carlo model used to simulate the oven behaviour is presented and the simulation results discussed. In the last section, simulation and experimental results are compared and discussed.

 \begin{figure}[ht!]
\begin{center}
   \includegraphics*[width=0.8\textwidth]{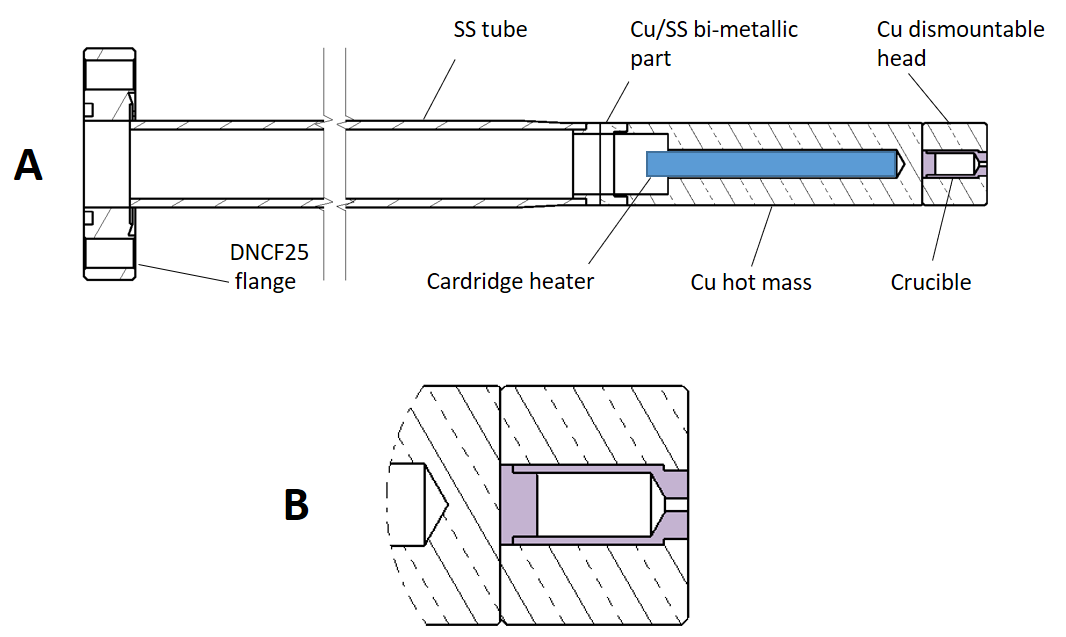}
    \caption{\small{(A) Calcium oven cut away view . (B) Detailed view of the crucible in light purple. }}
    \label{fig:fig1}
\end{center}
\end{figure}
 
  \section{Calcium oven}

The oven design is derived from an existing technology developed at Lawrence Berkeley National Laboratory \cite{CA_OVEN_BERKELEY}. A cutaway view of the oven is presented on  Fig.\ref{fig:fig1} with information on its mechanical composition.  The oven crucible is made of molybdenum to prevent chemical reaction with the metallic sample. The crucible cavity has a symmetry of revolution and is composed of two parts (see view B of Fig.\ref{fig:fig1}): (a) a 5mm diameter and 11 mm long cylindrical container ending on the last millimeter by a 30° cone  (shape imposed by the drilling tool geometry), (b) an 1 mm diameter and 2 mm long extraction channel (also referenced later as a nozzle). The oven  is heatable up to 875K when a Joule power of 200W is applied. A thermal simulation done with Ansys software has shown that the crucible temperature is very homogeneous with a maximum temperature gradient of the order of a few degrees Kelvin only.
  
\begin{figure}[h!]
\begin{center}
    \includegraphics*[width=0.8\textwidth]{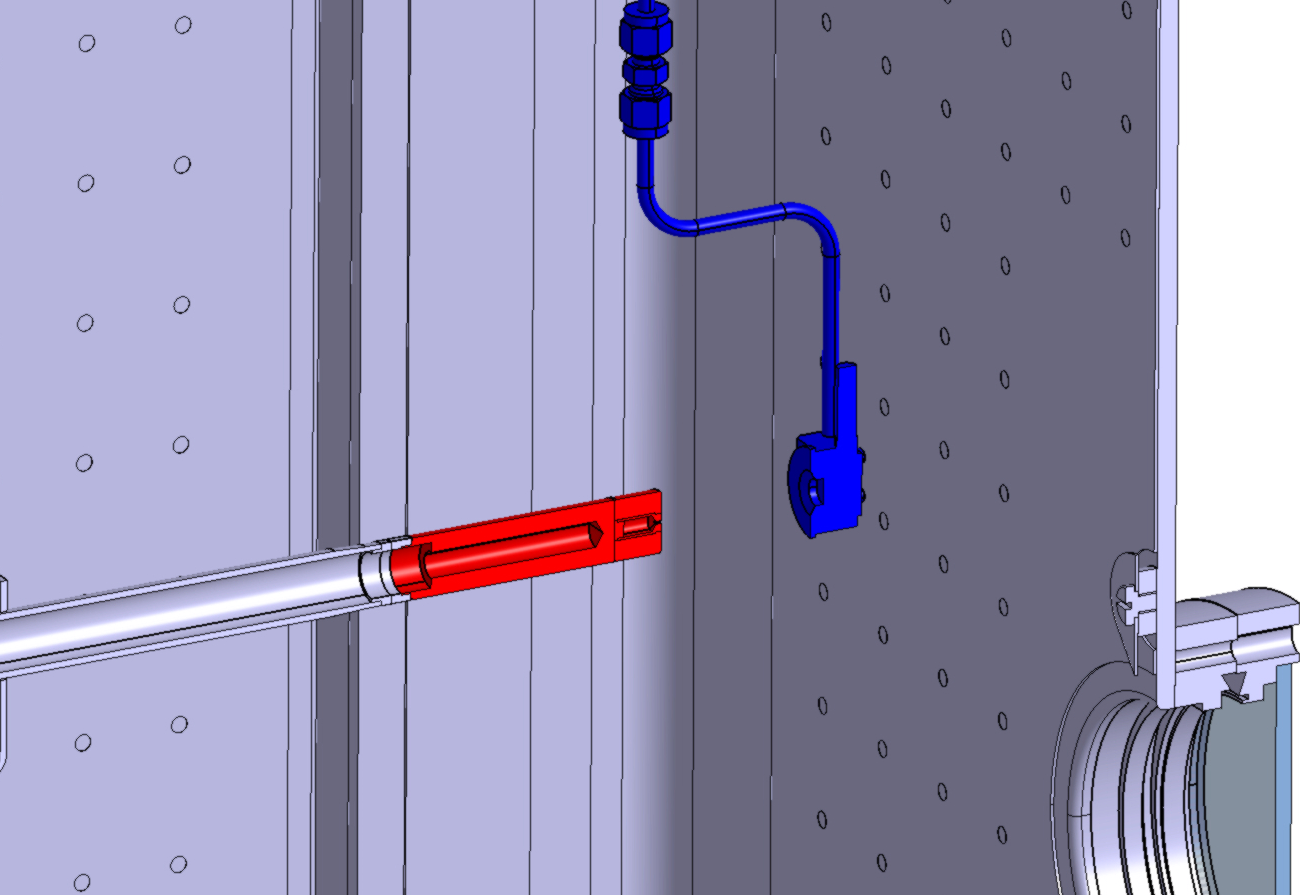}
    \caption{\small{Cutaway view of the vacuum chamber, the rotatable quartz measurement system (blue) and the oven tested (red) }}
    \label{fig:fig2}
\end{center}
\end{figure}
\section{Experimental measurements}
\subsection{Setup}
\begin{figure}[h!]
\begin{center}
    \includegraphics*[width=0.8\textwidth]{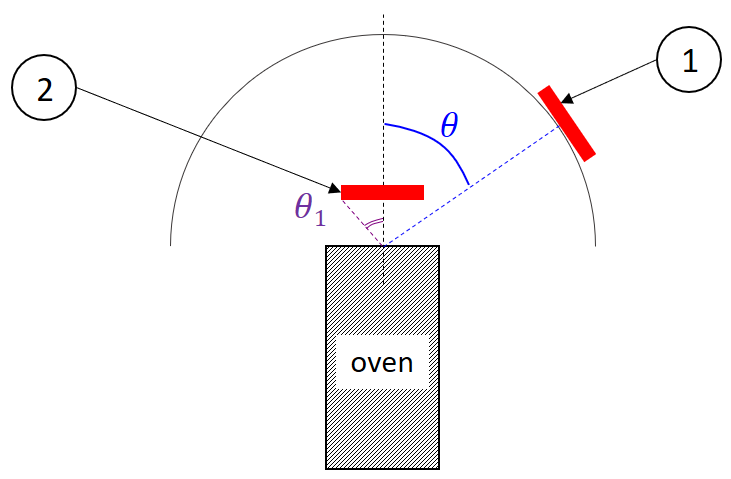}
    \caption{Sketch showing the positions of the quartz balance during the experiments.  (1) The quartz describes a circle of radius 60 mm around the oven exit. The quartz and the oven axis form an angle with $-\pi/2\le\theta\le\pi/2$. (2) The quartz is set  10 mm away from of the oven hole.}
    \label{fig:setup}
\end{center}
\end{figure}

The oven metallic atom  emission has been measured in  a dedicated vacuum chamber (see Fig. \ref{fig:fig2}) with a residual pressure $P_{0}=10^{-7}$ mbar. The oven temperature is monitored by the thermocouple included in the oven heater cartridge. The atom flux is measured with a quartz AUDA6 Neyco micro balance. The quartz is inserted into a  mechanical support resulting in an active measurement disc diameter of 8.1 mm. The quartz temperature is fixed thanks to  a water cooling system. The vertical support is mechanically connected to a rotatable vacuum flange by means of a tube bringing water cooling to the balance. the tube is bent to form two 90° bends so that during a rotation, the quartz describes a circle with a  60 mm radius. The crystal frequency of vibration is  ~3 MHz when a static electric field is applied. When evaporated metallic atoms are deposited on the quartz surface, its mass increases and thus changes its mechanical resonance frequency. The quartz frequency measurement is done with a dedicated Inficon controller which displays  the instantaneous frequency and calculates the mass per cm$^2$ accumulated during a programmable integration time. During the experiments, the mass flux is calculated after an integration time from 60 to 180 s. In our experimental conditions, The mass limit accuracy of the system has been estimated to be $\approx0.5 ng/cm^{2}/s$. The oven axis is horizontal and set perpendicular to the balance axis of rotation. The oven position can be translated along the direction of its axis of revolution. The two experimental configurations reported are displayed on Fig.\ref{fig:setup}. In the configuration (1), the oven exit is placed on the balance axis of rotation. During the quartz rotation, the distance between the oven and the quartz is constant and equals to 60 mm. The angle between the quartz surface and the oven surface is noted $\theta$. In the configuration (2), the quartz is set 10 mm away from the oven exit. The solid angle covered by the quartz is then $\approx$0.459 sr, with a maximum angle of detection $\theta_{1}=22,0$°.

\begin{table}[h!]
\caption{Mass flow and FWHM measurements from protocol (2) (see Fig.\ref{fig:setup}).}
\label{tab:measurement0}
\begin{center}
\begin{tabular}{c c c c }
 element  & T (K) & $\dot{M} (ng/s)$ & FWHM (deg.)  \\ 
   \hline
   \hline
 Ca & 848  & 148.9$\pm$67 & 53.7 $\pm$7.3\\  
 \hline
  Ca  & 873 & 352.1$\pm$67 & 55.5 $\pm$7.2 \\  
 \hline
  Ca  & 898 & 649.3$\pm$67 & 62.3 $\pm$4.2 \\  
  \hline
\end{tabular}

\end{center}
\end{table}

\subsection{Measurements}

\begin{figure}
\centering
  \includegraphics[width=0.8\linewidth]{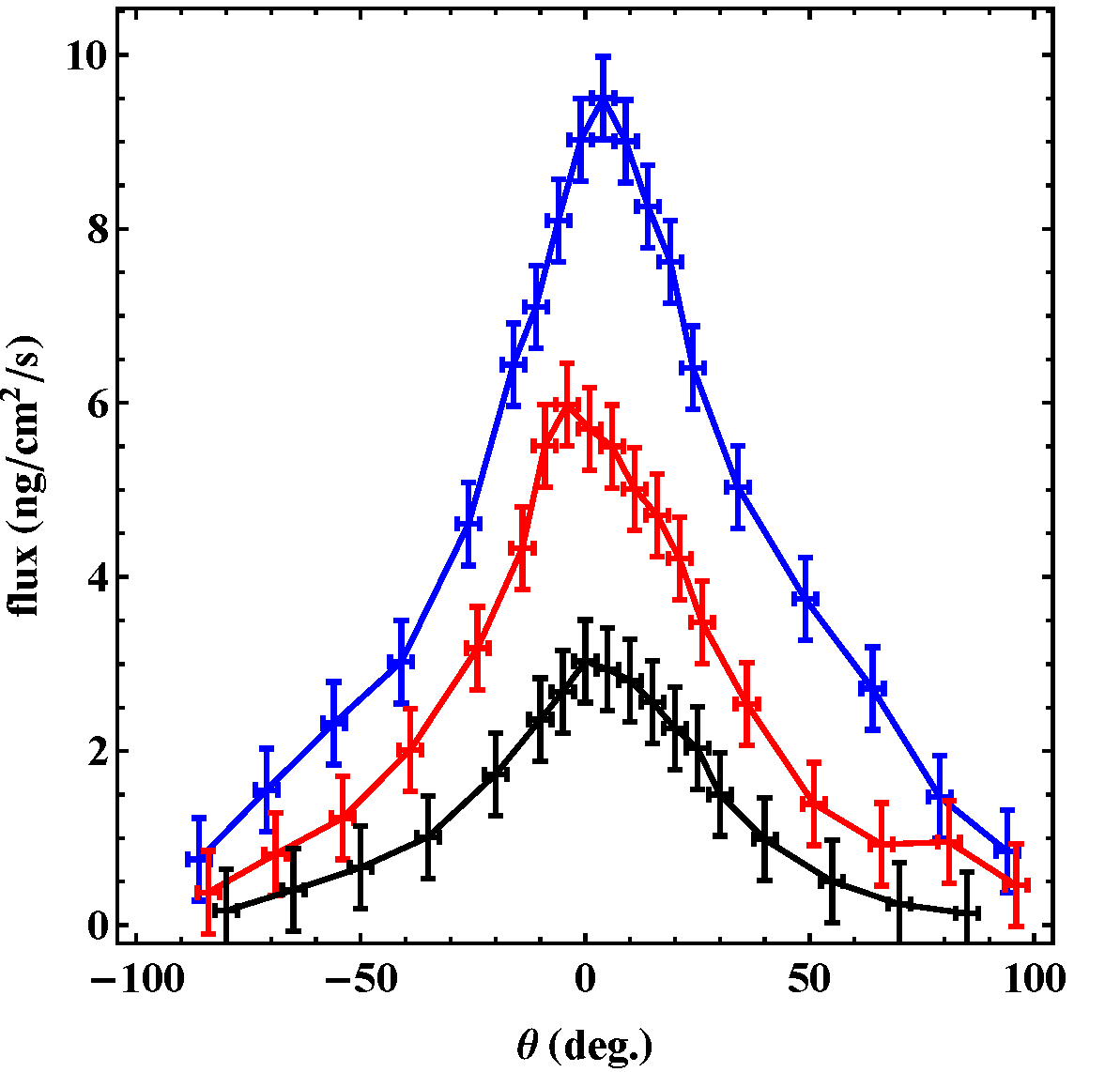}
\caption{Experimental flux emitted from the oven as a function of the angle $\theta$ for calcium. The black, red and blue plots are respectively measured at the temperatures  850, 875, 900°K.}
\label{fig:fig4}
\end{figure}

The oven run presented used a calcium 40 sample weighting 0.0909 g with a surface $s_{Ca}=0.8  \pm  0.2 \: cm^2$. The overall crucible internal surface is $s_c=2.8 \: cm^2$. The differential metallic mass flow emitted by the oven:

\begin{equation} \label{eq:solidangle0}
f(\theta)=\frac{1}{r^2}\frac{d\dot{m}}{ d\omega}
\end{equation}

 was measured  as a function of $\theta$ at T=848, 873 and 898 K (see protocol (1) in Fig.\ref{fig:setup}). Here $r$ is the radial distance between the oven and the balance and $d\omega$ stands for the differential solid angle taken at $\theta$. Results are displayed in fig.\ref{fig:fig4}. The errorbar on  $\theta$  is due to the extended surface of detection ($\sigma_\theta=0.9$°), the error on angle measurement ($\sigma_\theta\approx 2$°) and the error on alignment ($\sigma_\theta\approx 1.$°), leading to a global error $\sigma_\theta\approx2.4$°. The mass flux measurement precision is limited by the resolution of the balance for the chosen time of  integration of 180 s. The mass flux error is thus estimated to be $\sigma_{\dot{m}}\approx 0.5 ng/cm^2/s$. The full width half maximum (FWHM) of the distributions are reported in Table \ref{tab:measurement1}, along with the reconstructed total mass flux $\dot{M}$ integrated over $2\pi$ sr (and opportunely averaged on the overabundant range of measurement from $-\pi/2$ to $\pi/2$) :

\begin{equation} \label{eq:solidangle}
\dot{M}=2\pi r^2\left(\frac{1}{2}\int_{0}^{\pi/2} f(\theta) sin\theta d\theta+\frac{1}{2}\int_{0}^{-\pi/2} f(\theta) sin\theta d\theta\right)
\end{equation}

\begin{figure}[h!]
\begin{center}
    \includegraphics*[width=0.8\textwidth]{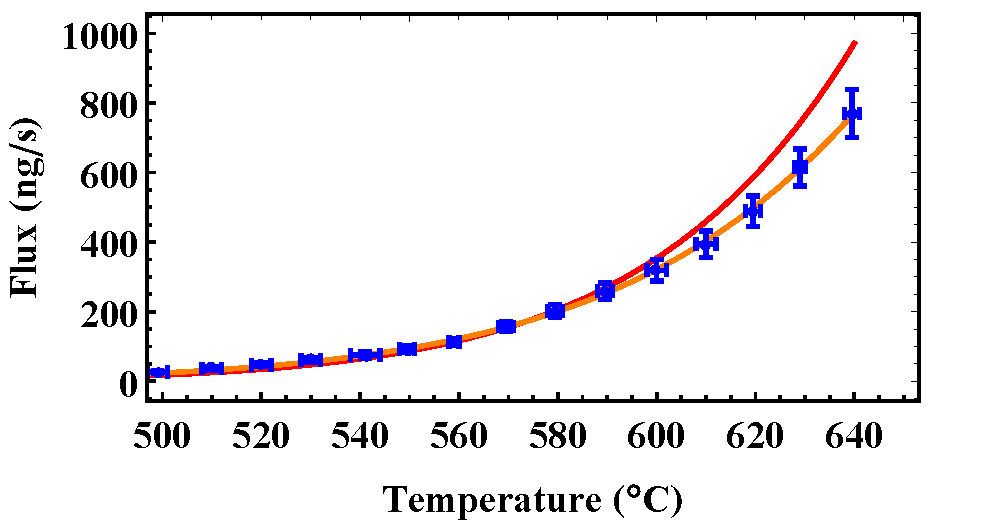}
    \caption{Total  calcium flux from the oven as a function of the temperature. Blue symbols: experimental data. Red: theoretical data with $S=s+s_c$ and Ca data from \cite{antoine}. Orange: least mean square fit of experimental data using $S=s+s_c$ and Antoine's law (see. Eq. \ref{eq:antoine}) coefficients $A$ and $B$ as fit parameters}
    \label{fig:massvsT}
\end{center}
\end{figure}

The angular FWHM values are consistent for all data (FWHM~53-63°). Further analysis of the angular mass flow distribution is proposed later in the text helped with a Monte Carlo code. Next, the calcium evaporation rate was measured as a function of the oven temperature in the experimental condition (2) (see Fig.\ref{fig:setup}).  The total calcium flux is reconstructed as a function of the temperature using the experimental measurement of $f(\theta)$, by applying the following correcting factor:

\begin{equation} \label{eq:ratio}
\frac{\int_{0}^{\theta_{1}} f(\theta) sin\theta d\theta}{\int_{0}^{\pi/2} f(\theta) sin\theta d\theta}=5.565
\end{equation}

where $f$ is taken for T=898K, when the precision of experiment (1) is the highest. 
The experimental data is plotted with blue errorbars on fig. \ref{fig:massvsT}. The other solid color plots on fig. \ref{fig:massvsT} are discussed later in the text.

\section{Oven thermodynamics and analysis}

\begin{table}[h!]
\caption{Calcium Antoine's equation coefficient according to \cite{antoine} and calculated in this study, expressed in standard units.}

\begin{center}
\begin{tabular}{ c c c c}

   & A & B & C \\ 
   \hline
   \hline
 Ca from \cite{antoine}& $10.34$ & $8.94\times10^{3}$ & 0\\  
 \hline
 Ca this work& $8.98  \pm 0.07$ & $7.79 \pm 0.11\times10^{3}$ & 0\\  
 \hline
\end{tabular}
\label{tab:antoine}
\end{center}
\end{table}
The aperture hole surface of the crucible ($0.78\, mm^{2}$) is small compared to its internal total surface ($s_{c}=\,2.2 cm^{2}$). See fig.\ref{fig:setup}, view B to visualise the difference. The cavity is thus sufficiently closed to consider it as a Knudsen cell\cite{knudsencell1,knudsencell2}. Consequently, when the metal sample is heated, the local pressure in the crucible raises rapidly to reach the saturating vapor pressure, well above the residual pressure of the vacuum chamber ( $P_{0}=10^{-5}$ Pa). The calcium saturation vapor pressures $P$(Pa) follow the Antoine's law as a function of the temperature :
\begin{equation} \label{eq:antoine}
log_{10}P=A-\frac{B}{C+T}
\end{equation}
where $T(K)$ is the metal temperature  and $A,B,C$ are thermodynamics parameters, unique for each chemical element. 
The evaporated mass rate $\dot{M}$ emitted from the solid metal in the crucible can be expressed using the Hertz-Knudsen equation\cite{langmuir_1,langmuir_2} :

\begin{equation}\label{eq:hertzknusdenequation}
\dot{M}=P  \sqrt{\frac{m}{2 \pi kT}} S
\end{equation}

where $m$ is the atom mass, $P$ is the metal saturating vapor pressure and $S$ is the surface of evaporation. Another important oven parameter is the sticking time of atoms on the hot crucible surfaces defined by the Frenkel equation\cite{FRENKEL}: 

\begin{equation}\label{eq:frenkel}
    \tau=\tau_{0}e^{\frac{H}{kT}}
\end{equation}
where $\tau_{0} \approx 10^{-13} s$ is the atom vibration period on the surface \cite{ Glebovsky}, $H$ is the enthalpy of the adsorbed atom, $k$ the Boltzmann constant and $T$ the surface temperature. For the calcium case, two sticking times must be considered. the first is $\tau_{Ca-Ca}$, the sticking time of calcium atom on the metallic Ca surface sample. the other is $\tau_{Ca-Mo}$, the sticking time of Ca on Mo. The adsorbed Ca enthalpy on Ca is well documented : $H_{Ca}=1.6 eV$, giving a sticking time of  $1-1000 ms$ for $T~600-900 K$ . On the other hand, the adsorbed Ca on Mo enthalpy value $H_{Ca-Mo}$ was not found in the literature. Nevertheless, $H_{Ca-Mo}$ is expected to be much higher than $H_{Ca}$. A rough estimate of $H_{Ca-Mo}$ value can be considered from $H_{Sc-Mo}=5.5 eV$ available in \cite{ENTHALPY}. In the present experimental conditions, $\tau_{Ca-Ca}\ll \tau_{Ca-Mo}$ on all the temperature range covered. Hence, when the calcium evaporation starts, a layer of calcium should first be formed on the Mo crucible wall and stay stuck on Mo. Once the layer is complete, next Ca adsorption on the surface is done on Ca, which strongly reduces the sticking time and transforms the crucible surface to a fresh extra source of calcium. Two experimental confirmations of this phenomenon have been observed: (i) at first start, the calcium flux is very small for a given temperature and, after a sufficiently long time of operation (of the order of an hour), the flux amplifies to reach a much higher stable value. (ii)  Immediately after venting the oven used and inspecting its crucible, the presence of a Ca layer is visible on the crucible surface which very  rapidly get oxidised to form a white CaO powder. This hypothesis is further investigated helped with the experimental data from experiment (1) (see Fig.\ref{fig:setup}). 
 The theoretical mass flow expected from Eq.\ref{eq:antoine} and Eq.\ref{eq:hertzknusdenequation} using Ca data from \cite{antoine} is calculated for the temperatures 848, 873 and 898 K, considering the two evaporating surfaces $S=s$ and $S=s_c+s$ and reported in Tab.~\ref{tab:theorymassflow} along with the experimental mass flow from Tab.~\ref{tab:measurement0}. Clearly, one can see that the sole metal sample evaporating surface (column with $S=s$) is insufficient to reproduce the total mass rate from the oven. But when the crucible surface is added (column with $S=s_c+s$), the matching between theory and experiment becomes very close. The measurement confirms the hypothesis of the transient formation of a Ca layer on the crucible surface which, once completed enhances the evaporation proportionally to the crucible surface.
 
  \begin{table}[h!]
\caption{Theoretical integrated Mass flow from the oven in (ng/s) derived from Eq.\ref{eq:antoine} and \ref{eq:hertzknusdenequation}, compared with the experimental flux for T=848, 873 and 898K.}
\begin{center}
\begin{tabular}{c c c c c}
T& P & $\dot{M}(s)$ & $\dot{M}(s+s_c)$ & $\dot{M}$ Exp.\\
(°K) & (Pa) & ng/s & ng/s & ng/s\\ 
   \hline
   \hline
848 & 0.627 & 47.7  & 179& 148.9$\pm$67\\  
 \hline
873 & 1.26 & 94.5  &  354& 352.1$\pm$67\\
 \hline
898 & 2.42 & 179  &  671& 649.3$\pm$67\\
 \hline
\end{tabular}

\label{tab:theorymassflow}
\end{center}
\end{table}

Result of experiment in condition (2) (see Fig.\ref{fig:setup})  is next compared with the prediction of eq.\ref{eq:hertzknusdenequation} in the fig.\ref{fig:massvsT}. The sensitivity to temperature comes here through the pressure and the Antoine's eq.\ref{eq:antoine}. The experimental data is the blue errorbar, the mass flow prediction according to the  eq.\ref{eq:antoine} using the known tabulated coefficients for calcium, (see Tab.\ref{tab:antoine}), is plotted in solid red line. One can see a fairly good fit with experimental data up to 600°C (873K). Above this value, experimental flux is lower than the semi-empirical prediction. Because the total flux has been reconstructed with a sufficiently high precision, alternative Antoine's coefficients are proposed for calcium to fit the data resulting in the orange solid line. The fit value found for the whole temperature range are $A=8.978\pm 0.07$, $B=7787.5\pm 110$ and $C=0$ in standard units.

\section{Monte Carlo Simulation}\label{sec:MC}

\begin{figure}
\centering

  \includegraphics[width=0.8\linewidth]{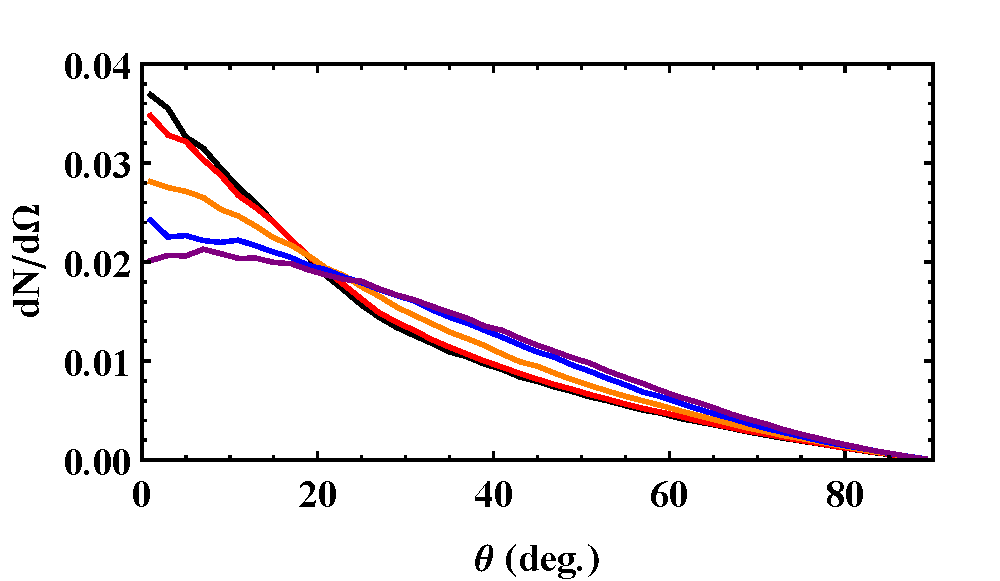}
 
\caption{Normalized differential distribution of atoms at the exit of the oven per unit of solid angle as a function of $\theta$ for different temperatures: 600K (black), 900K (red), 1000K (orange), 1100K (blue), 1200K (purple).}
\label{fig:dNdOmega}
\end{figure}

\begin{figure}
\centering

  \includegraphics[width=0.8\linewidth]{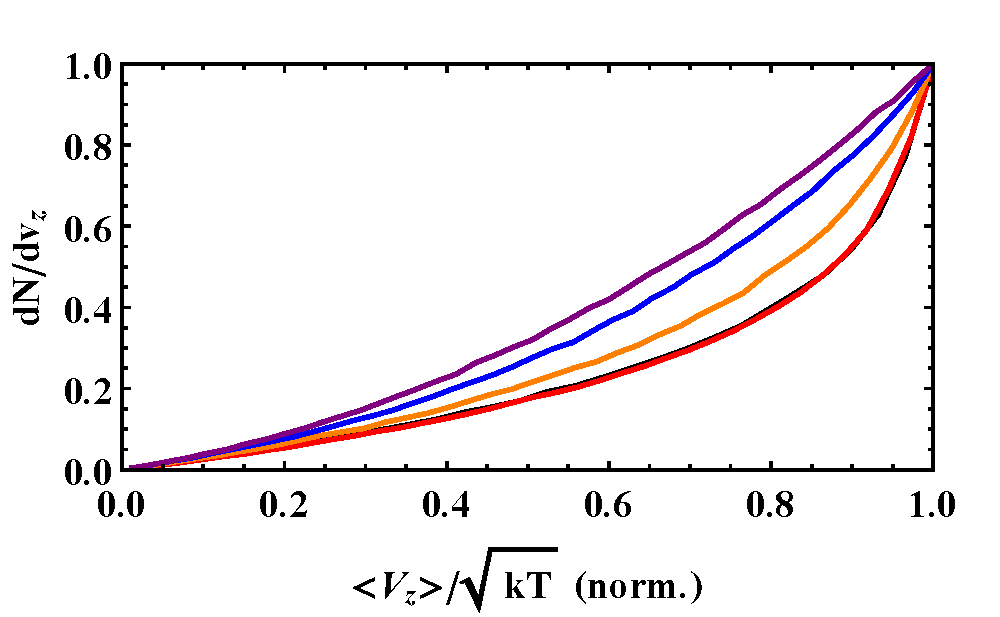}
 
\caption{Normalized distribution of atoms velocity at the exit of the oven along the direction z for different temperatures: 600K (black), 900K (red), 1000K (orange), 1100K (blue), 1200K (purple).}
\label{fig:dNdVz}
\end{figure}

\begin{figure}
\centering

  \includegraphics[width=0.8\linewidth]{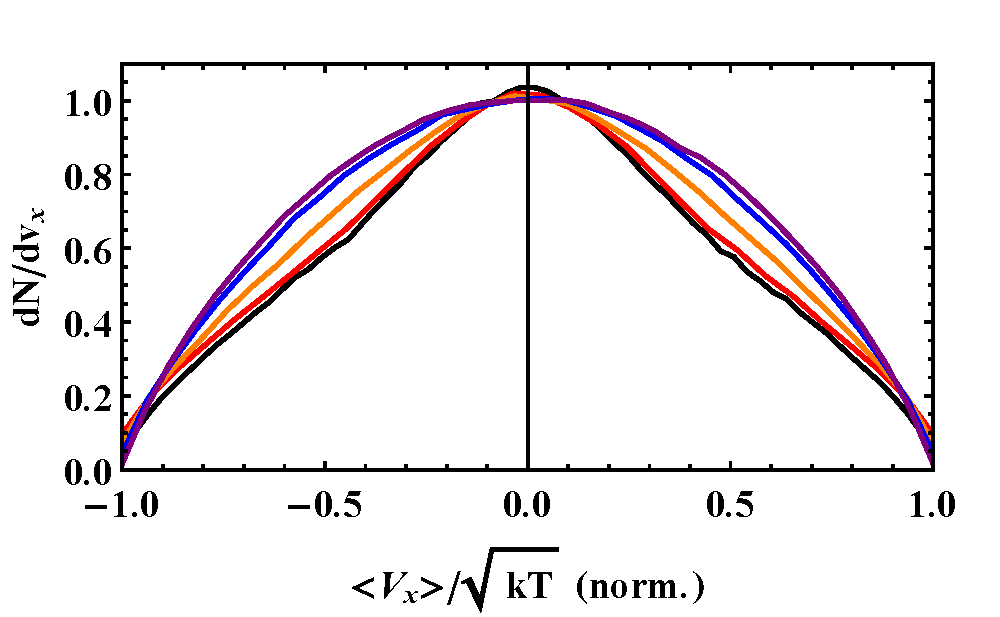}
 
\caption{Normalized distribution of atoms velocity at the exit of the oven along the direction x for different temperatures: 600K (black), 900K (red), 1000K (orange), 1100K (blue), 1200K (purple).}
\label{fig:dNdVx}
\end{figure}

The atom  angular distribution from the oven has been investigated by means of a Monte Carlo simulation and compared with the experimental data. The exact 3 dimensions crucible geometry is considered. The oven temperature $T$ is a free parameter and is assumed to be uniform over all the crucible. The metallic sample geometry is not modelled and the initial atom emission position is done randomly along a line following the bottom part of the crucible part (a). The atom emission from the wall follows the Lambert's cosine law:
\begin{equation}\label{eq:cosinelaw}
    P(\theta)=cos\theta
\end{equation}
where $P(\theta)$ is the probability of emission at an angle $\theta$ with respect to the local normal to the wall. A special care must be taken to inverse this probability distribution function appropriately to use it safely in the Monte-Carlo code\cite{cosinelaw}. Because the atom emission from the wall is done in a cavity with convex walls, a test is added in the code to check if a fresh reemission occurs toward the cavity and not to the wall. The atom velocity is taken as the mean thermal velocity:
\begin{equation}\label{eq:velocity}
    v=\sqrt{\frac{3kT}{m}}
\end{equation}
where $m$ is the considered atom mass. No accomodation from the wall is considered as particles are assumed to be isothermal. Each new adsorption at the wall is counted and re-emission is immediate using the cosine law. The pressure in the part (a) of the crucible is considered constant at the saturating vapor pressure $P$ of eq.\ref{eq:antoine}. The pressure in the extraction channel (b) is assumed to decrease linearly from $P$ down to the vacuum chamber pressure $P_0=10^{-5}$ Pa. The atom collisions are modelled in the crucible using the atom mean free path $\lambda$:

\begin{equation}\label{MFP}
    \lambda=\frac{kT}{\sqrt{2} \pi l^{2}P}
\end{equation}

where $l$ is the atom diameter ($l\approx 360 pm$ for calcium). At each time step, the local pressure is considered and a random number is generated to check whether the atom collides or not. In case of collision, a new atom velocity direction is randomly generated on a uniform sphere. The metallic gas regime in the oven can be assessed with the Knudsen number:

\begin{equation}
    K=\frac{\lambda}{d}
\end{equation}\label{eq:knudsennumber}

where $d$ is a characteristic length of the system studied. When $K>0.5$, the gas is collisionless and atoms exiting the oven are directly those emitted from the walls. An intermediate regime occurs when $0.01<K<0.5$ where the volume collisions start to play a role and finally, when $K<0.01$, the gas becomes fully collisional and the effect from the wall is secondary. In our case, the crucible has 2 characteristic lengths and hence two Knudsen numbers: $ K_{i}=\frac{\lambda}{d_{1}}$ with $d_{1}$=5 mm for the main cylinder and  $ K_{ii}=\frac{\lambda}{d_{2}}$ with $d_{2}$=1 mm for the crucible exiting channel. In this latter case, we considered the mean pressure in the exit channel $\frac{1}{2}(P+P_{0})$ to calculate $\lambda$. The evolutions of $\lambda$, $K_{i}$ and $K_{ii}$ are proposed as a function of $T$ and $P$ in the table \ref{tab:MFP}. For the measurements performed on calcium up to 900K, one can deduce that the calcium gas behaviour is mainly non-collisional. Nevertheless, for T=900K, $K_{i}=0.63$ is at the threshold and collision effect, thought not dominant should start to play a role. At 1000K, the calcium gas is collisional in the bulk area (a) while it remains above the transition in the exiting channel. a fully developed collision regime is reached at 1100K with $K_{ii}=0.12$. The differential  angular distribution of exiting atom per unit of solid angle is displayed in Fig.\ref{fig:dNdOmega} as a function of the temperature. The number of particles generated to produce the curves is 1 million. One can clearly see the transition from the non collisional regime at 600K (see black curve) to the intermediate regime at 1000K where the bulk crucible is collisional while the exit channel is not (see orange curve) and finally above 1100K where the whole oven volume is collisional and the extraction emittance is finally defined by the sole exiting channel geometry (see blue and purple curves). The associated normalized distribution of  axial and radial exiting atom velocities are reported on Fig.~\ref{fig:dNdVz} and ~\ref{fig:dNdVx} respectively as a function of the oven temperature. The  convention of color is identical as for Fig.\ref{fig:dNdOmega}. One can note how the exit channel geometry favors the atom emission in the non-collisional regime ~20° around the z direction (oven axis), while this condition is destroyed above a temperature of 1000K.

\begin{table}[h!]
\caption{Evolution of the atom mean free path $\lambda$ and the two  Knudsen numbers $K_{i}$, $K_{ii}$ for the calcium oven as a function of the temperature $T$ and the saturating vapor pressure $P$.}
\begin{center}
\begin{tabular}{c c c c c}
  T  & P &$\lambda$ & $K_{i}$ & $K_{ii}$\\
  (K) &(Pa) & (m) &  & \\ 
   \hline
   \hline
 600 & $2.75\times10^{-5}$ & 194.4 & 38895  & 285348\\  
 \hline
 700 & $3.70\times10^{-3}$ & 1.68  &  337 & 3365\\  
 \hline
 800 & 0.146 & 0.0488  &  9.76 & 97\\  
 \hline
 850 & 0.66 & 0.011 & 2.2846 & 22.8\\
 \hline
 900 & 2.55 & 0.0031  & 0.63 & 6.3\\  
 \hline
 1000 & 25.19 & $3.55\times10^{-4}$  &  0.071 & 0.71\\  
  \hline
  1100 & 163.2 & $6.01\times10^{-5}$  & 0.012 & 0.12\\  
  \hline
\end{tabular}
\label{tab:MFP}
\end{center}
\end{table}

\begin{table}[h!]
\caption{Evolution of the mean number of atom bounces, volume collision and distance travelled by an atom before its extraction from the oven as a function of the temperature.}
\begin{center}
\begin{tabular}{c c c c }
  T (K) & bounces & collisions & distance (m) \\
   \hline
   \hline
 600 & 687 & 0.01 & 2.75  \\  
 \hline
 700 & 686 & 2.03  & 2.77 \\  
 \hline
 800 & 690 & 55.8 & 2.80  \\  
 \hline
 850 & 769 & 245 & 2.84 \\
 \hline
 900 & 816 & 998  & 3.21  \\  
 \hline
 1000 & 1699 & 18416 & 7.02 \\  
  \hline
  1100 & 3291 & 2.89$\times 10^{5}$  & 24.14 \\  
  \hline
\end{tabular}
\label{tab:SIMUMRESULTS}
\end{center}
\end{table}

\begin{figure}
\centering
\includegraphics[width=0.8\linewidth]{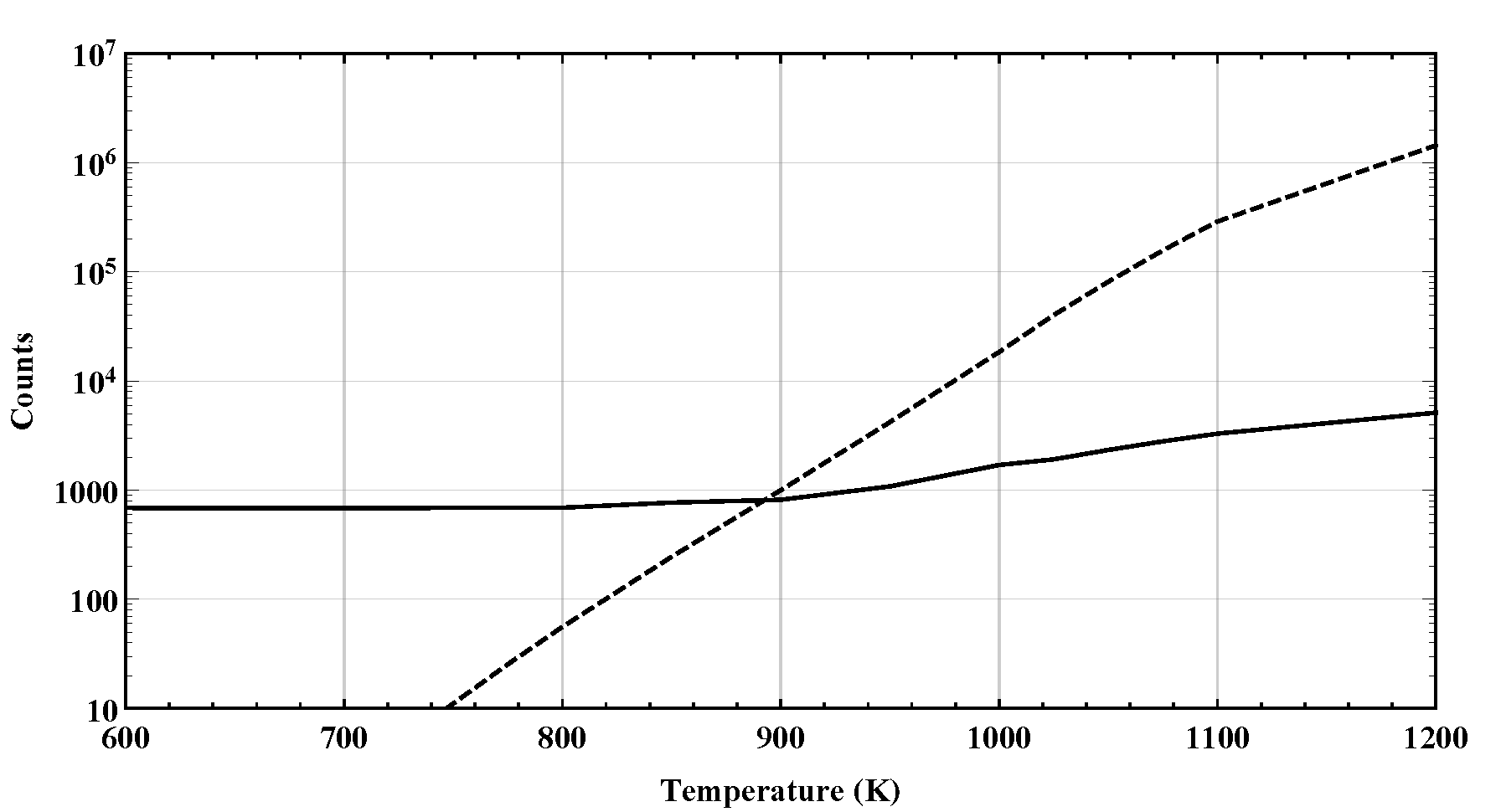}
\caption{(Black) Mean number of atom adsoption on the crucible wall before atom extraction. (Dashed Black) Number of volume collision in the oven before extraction.}
\label{fig:counts}
\end{figure}

\begin{figure}
\centering
\includegraphics[width=0.8\linewidth]{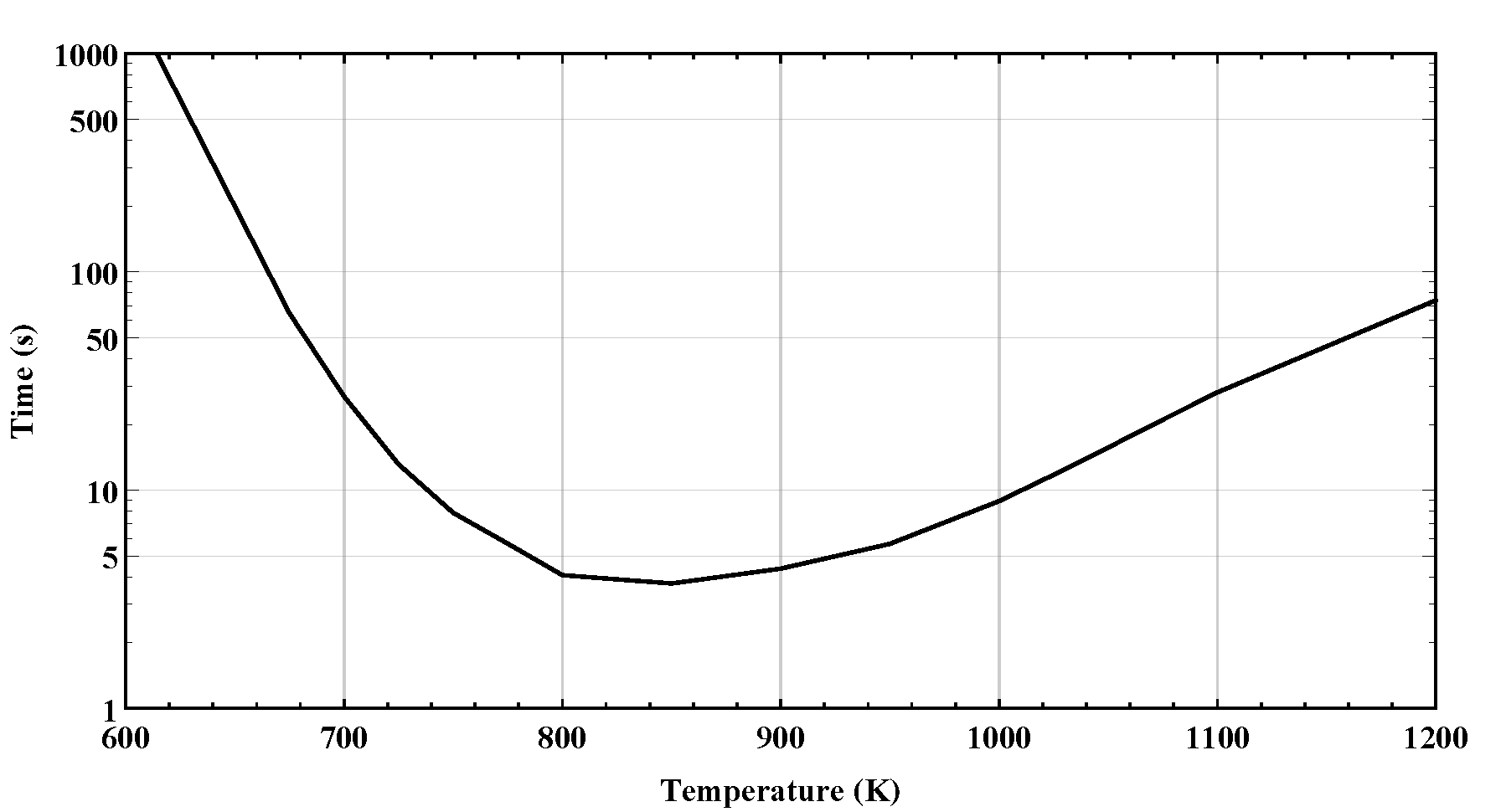}
\caption{Estimated mean time to extract an atom from the oven as a function of temperature.}
\label{fig:time}
\end{figure}

Table \ref{tab:SIMUMRESULTS} presents the evolution of the mean number of atom bounce on the crucible surface, the mean number of atom volume collision and the mean distance travelled before the atom extraction from the oven. Figure \ref{fig:counts} shows the evolution of the mean numbers of bounce and collision versus the temperature. The temperature at which the volume collision becomes dominant is 900K. It is worth noting that the mean number of bounce increases much more slowly with temperature than the number of volume collision. The evolution of the mean extraction time as a function of the temperature is estimated using the following formula:
\begin{equation}\label{eq:extractiontime}
    t=d/v+b\tau_{0}e^{\frac{H}{kT}}
\end{equation}
where $b$ is the mean bounce number, $v$ and $d$ the mean atom velocity and distance travelled. The result is displayed in Fig.\ref{fig:time}. The time of extraction first decreases rapidly between 600 to 850 K thanks to the reduction of the mean sticking time. Above 850 K, the extraction time increases again as now the process is dominated by the mean distance travelled which increases faster than the mean atom velocity.

\begin{figure}[h!t]
\centering
\includegraphics[width=0.8\linewidth]{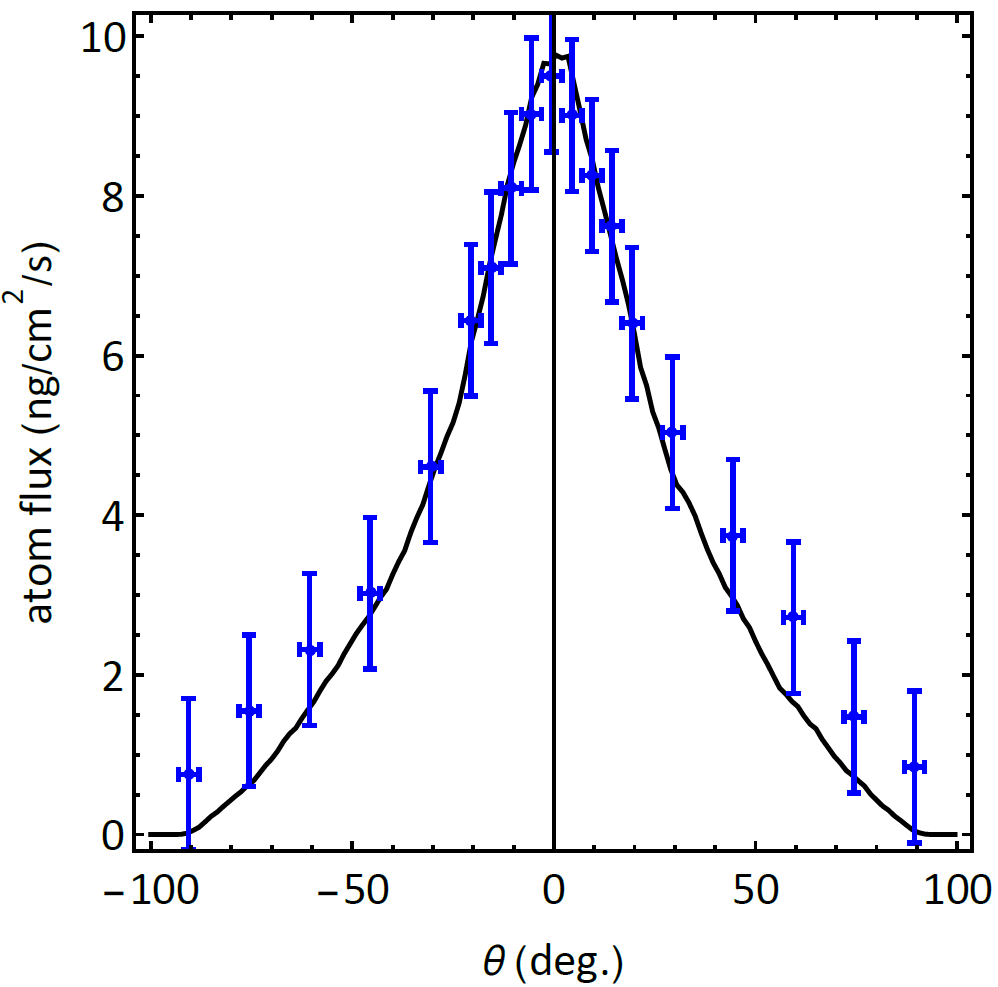}
\caption{Comparison between the experimental differential mass flux measured in the micro-scale and the Monte-Carlo simulation at T=898K.}
\label{fig:comp_simu_exp}
\end{figure}

\begin{table}[h!t]
\caption{Comparison between FWHM (in degree) of angular differential mass flux measurements and simulation for calcium.}
\begin{center}
\begin{tabular}{c c c }

  T (K) & exp. (°) & simulation (°)  \\ 
   \hline
   \hline
 848   & 53.7 $\pm$7 & 48.6$\pm$ 0.6\\  
 \hline
  873  & 55.5 $\pm$7 & 51.4$\pm$ 0.6 \\  
 \hline
  898  & 62.3 $\pm$7 & 54.45$\pm$ 0.6 \\  
  \hline
  \hline
\end{tabular}
\label{tab:measurement1}
\end{center}
\end{table}

\section{Comparison of simulation and experiment }\label{sec:COMP}

The oven Monte Carlo code is used to simulate the theoretical atom flux expected on the micro scale as a function of the angle $\theta$ (see fig.\ref{fig:setup}). An example of the comparison between simulation and measurement is proposed on Fig.\ref{fig:comp_simu_exp}. The simulation reproduces well the general shape of the differential flux as a function of the angle $\theta$. A discrepancy is nevertheless visible for angle larger than 45° where the measurements are higher than the simulation. the difference is nevertheless included in the error bar. The differential flux FWHM is calculated for the simulation and compared with the experimental measurements. The results of the simulation are consistent with the measurements. the experimental FWHM is always higher than the simulation. This systematic difference is likely due to a non simulated physical effect. One could suggest for instance that the actual oven extraction geometry is not exactly as simulated, or that the atom flux continues to collide at the exit of the oven on the first mm. As a conclusion,  The Monte Carlo code developed provides satisfactory results compard to experimental measurements and will next be used to study the metallic atom capture in ECR ion source plasma.
\newpage
\appendix
\section{Experimental data}

The reconstructed experimental values of the total atom mass flux plotted in fig \ref{fig:massvsT} are reported for convenience in the table\ref{tab:measurement2}.

\begin{table}[h!]
\caption{total calcium mass flow as a function of the temperature for the configuration (2).}
\begin{center}
\begin{tabular}{c c}
 T (k)  & $\dot{M}$ (ng/s)   \\ 
   \hline
   \hline
 489.5$\pm$1.5&19.0$\pm$0.10\\
 499$\pm$2&27.3$\pm$0.11\\
 510$\pm$2& 36.6$\pm$0.11\\
 520$\pm$2&48.$\pm$0.11\\
 530$\pm$2&61.$\pm$0.10\\
 541$\pm$3&75.7$\pm$0.12\\
 549.5$\pm$1.5&92.2$\pm$0.09\\ 
 559$\pm$1&113.7$\pm$0.09\\
 569.5$\pm$1.5&157.2$\pm$0.09\\
 579.5$\pm$1.5& 202.6$\pm$0.09\\
 589.5$\pm$1.5&258.8$\pm$0.10\\ 
 600$\pm$2&319.1$\pm$0.10 \\
610$\pm$2&394.3$\pm$0.09\\
619.5$\pm$1.5&488.4$\pm$0.09\\
629$\pm$1&615.$\pm$0.09\\
639.5$\pm$1.5&769.7$\pm$0.09\\
  \hline
\end{tabular}
\label{tab:measurement2}
\end{center}
\end{table}

\end{document}